\documentclass[12pt]{article}
\usepackage{graphicx}
\usepackage{epstopdf}
\usepackage{dcolumn}
\usepackage{verbatim}
\usepackage{float}
\usepackage{slashed}
\usepackage[hidelinks]{hyperref}

\newcommand\pubnumber{DPF2015-22}
\newcommand\pubdate{\today}

\def\mit{Laboratory for Nuclear Science\\
Massachusetts Institute of Technology\\77 Massachusetts Avenue, Bldg 26-568, MA 02139, U.S.A}
\def\support1{}

\def\msstate{Department of Physics and Astronomy\\
Mississippi State University\\P. O. Box 5167, Mississippi State, MS 39762, U.S.A}
\def\support2{}

\def\Title#1{\begin{center} {\Large #1 } \end{center}}
\def\Author#1{\begin{center}{ \sc #1} \end{center}}
\def\Address#1{\begin{center}{ \it #1} \end{center}}

\newcommand\pubblock{\rightline{\begin{tabular}{l} \pubnumber\\
         \pubdate  \end{tabular}}}
\newenvironment{Abstract}{\begin{quotation}  }{\end{quotation}}
\newenvironment{Presented}{\begin{quotation} \begin{center} 
             PRESENTED AT\end{center}\bigskip 
      \begin{center}\begin{large}}{\end{large}\end{center} \end{quotation}}

\def\beq{\begin{equation}}
\def\eeq#1{\label{#1}\end{equation}}
\def\eeqn{\end{equation}}

\def\beqa{\begin{eqnarray}}
\def\eeqa#1{\label{#1}\end{eqnarray}}
\def\eeqan{\end{eqnarray}}

\let\bar=\overbar

\def\Dslash{\not{\hbox{\kern-4pt $D$}}}
\def\dslash{\not{\hbox{\kern-2pt $\del$}}}

\def\msb{{\bar{\ssstyle M \kern -1pt S}}}

\begin{document}
\begin{titlepage}
\pubblock

\vfill
\Title{Test of Lorentz Invarience from Compton Scattering}
\vfill
\Author{ Prajwal Mohanmurthy \footnote{prajwal@mohanmurthy.com . Now at ETH Zurich \& Paul Scherrer Institute.}}
\Address{\mit}
\Author{ Dipangkar Dutta, Amrendra Narayan \footnote{Now at Indian Institute of Technology, Bombay.}}
\Address{\msstate}
\vfill
\begin{Abstract}
In the recent times, test of Lorentz Invariance has been used as a means to probe theories of physics beyond the standard model. We describe a simple way of utilizing the polarimeters, which are a critical beam instrument at precision and intensity frontier nuclear physics labs such as the erstwhile Stanford Linear Accelerator Center (SLAC) and Jefferson Lab (JLab), to constrain the dependence of vacuum dispersion with the energy of the photons and its direction of propagation at unprecedented level of precision. We obtain a limit of minimal Standard Model extension (MSME) parameters: $\sqrt{\kappa_X^2 + \kappa_Y^2} <  4.6 \times 10^{-10}$ and $\sqrt{\left( 2c_{TX} - (\tilde{\kappa}_{0^+}^{YZ} \right)^2 + \left( 2c_{TY} - (\tilde{\kappa}_{0^+}^{ZX} \right)^2} < 4.6 \times 10^{-10}$. We also obtain a leading constraint for the refractive index of free space $n = 1 + (2.44\times10^{-9} \pm 6.82\times 10^{-9})$.
\end{Abstract}
\vfill
\begin{Presented}
DPF 2015\\
The Meeting of the American Physical Society\\
Division of Particles and Fields\\
Ann Arbor, Michigan, August 4--8, 2015\\
\end{Presented}
\vfill
\end{titlepage}
\def\thefootnote{\fnsymbol{footnote}}
\setcounter{footnote}{0}
\section{Introduction}
Lorentz invariance was first introduced to explain the constant speed of light in all reference frames \cite{[0]}, but has formed the corner stone of modern standard model. Testing the Lorentz invariance rigorously is thus a very interesting research activity. There are direct tests of Lorentz invariance mostly involving studying the dependence of vacuum dispersion on a number of other physical properties such as its energy. According to the CPT theorem, the joint Charge-Parity-Time symmetry has to hold for all processes governed by a Lorentz invariant theory and vice-versa \cite{[2]}. Therefore, in particle and nuclear physics probing violations of CPT symmetry has been a means to test Lorentz invariance. Nuclear and particle physics methods often involve indirect tests of Lorentz invariance by measuring violations of CPT symmetry either by studying the joint CPT symmetry or by studying a subset of the CPT symmetry, such as CP symmetry violation or T symmetry violation independently.

Certain beyond the standard model (BSM) theories have been known to break CPT symmetry and Lorentz symmetry by extension. Ref. \cite{[3]} lists all known BSM theories that break CPT and Lorentz symmetry to date, while Ref. \cite{[4]} lists all the experimental limits on violation of CPT and Lorentz symmetry to date. The BSM theories define and make use of additional parameters currently not in the standard model and often these new parameters are called BSE parameters.

Compton scattering, which is currently the only way of continuously monitoring electron beam polarization precisely has spawned Compton polarimeters at each of the labs using polarized electron beam to probe nuclear matter such as SLAC, JLab, LEP, DESY, NIKHEF and MIT-Bates. Compton polarimeters have been demonstrated to achieve precision better than 1\% \cite{[5]}. The high degree of precision of Compton polarimeters allows precise measurement of vacuum dispersion competitive with the current leading limits at the photon energy of the back scattered photon. The continuous operation of these polarimeter over long periods allows investigation of the sidereal variation of the vacuum dispersion. The measured values of vacuum dispersion can also be used to obtain upper limits on the SME parameters.

\section{Method}
\subsection{Determination of refractive index of free space}
Usually the variation in vacuum dispersion is studied as a variation in the refractive index of vacuum. If the vacuum dispersion was constant w.r.t to a varying parameter, such as photon energy, the refractive index of vacuum is 1. For Compton scattering of electrons with initial energy $\epsilon_0$ and mass $m_e$, on photons with initial and final energy and angle $\omega_0, \theta_0$ and $\omega, \theta$ respectively, the Compton scattering cross section and longitudinal asymmetry are given by~\cite{ccref}:\\
\begin{equation}
\frac{d\sigma}{d\rho} = 2\pi r_e^2 a \left[ \frac{\rho^2(1-a)^2}{1-\rho(1-a)} + 1 + \left(\frac{1-\rho(1+a)}{1-\rho(1-a)}\right)^2\right]
\label{eq:cc}
\end{equation}

\begin{equation}
A_l(\rho) = \frac{2\pi r_e^2 a}{d\sigma/d\rho}(1-\rho(1+a)) \left[1 - \frac{1}{(1-\rho(1-a))^2}\right]
\label{eq:al}
\end{equation}

Where $r_e = \alpha \hbar c/m_ec^2 = 2.817\times10^{-15}$~m, is the classical electron radius, $\rho = \omega/\omega_{max}$ is the scattered photon energy normalized to its maximum value, and $a = \frac{1}{1+ 4\omega_0\epsilon_0/m_e^2}$ is a kinematic parameter. As demonstrated in Ref.~\cite{vahplb}, Compton scattering is very sensitive to tiny deviations of the refractive index from unity due to an amplification by the square of the initial Lorentz boost ($\gamma_0$). For photons scattering off ultra-relativistic electrons in vacuum with $n\approx 1$ (up to $\mathcal{O}[(n-1)^2]$), energy-momentum conservation gives~\cite{vahprl};

\begin{equation}
\epsilon_0x - \omega(1+x+\gamma_0^2 \theta^2) + 2\omega_0(1 - \frac{\omega}{\epsilon_0})\gamma_0^2(n-1) =0, 
\label{consveq}
\end{equation}
where $x = 4\gamma_0 \omega_0 \sin^{2}{(\frac{\theta_0^2}{2})}/m_e$. 

In the Compton polarimeter, the scattered electrons are momentum analyzed by a dipole magnet and detected on a position sensitive detector which measures the deflection of the scattered electrons with respect to the unscattered electrons. If the refractive index is independent of $\omega$, the effect of any deviation of the refractive index from unity can be incorporated into the Compton cross section and longitudinal asymmetry by modifying the scattered photon energy normalized to its maximum value, $\rho \rightarrow \rho(n)$ as \cite{vahplb}, 
\begin{equation}
\rho(n) = \rho \left[\frac{1 + \frac{2\gamma_0^2(n-1)(1+\gamma_0^2\theta^2)}{(1+x+\gamma_0^2\theta^2)^2}}{1 + \frac{2\gamma_0^2(n-1)}{(1+x)^2}}\right] \approx \rho \left[1 + 2\gamma_0^2(n-1)f(x,\theta)\right] + \mathcal{O}[(n-1)^2]
\label{eq:rhon}
\end{equation}
where the kinematic function $f(x,\theta) = ((1+\gamma_0^2\theta^2)(1+x)^2 - (1+x^2+\gamma_0^2 \theta^2)^2) / ((1+x+\gamma_0^2\theta^2)^2(1+x)^2)$.

The longitudinal asymmetry can then be re-written to incorporate the deviation of the refractive index from $n=1$ by replacing $\rho$ by $\rho(n)$ in Eq.~\ref{eq:al}. The new modified asymmetry is then fit to the measured asymmetry with three parameters; the product of electron beam and laser polarizations, $P_eP_{\gamma}$, the strip location of the Compton edge, $N_{CE}$ and $2\gamma_0^2(n-1)$. The new parameter $2\gamma_0^2(n-1)$ is used to extract the deviation of the refractive index from unity.

\begin{figure}[ht]
\centering
    \includegraphics[scale=0.4]{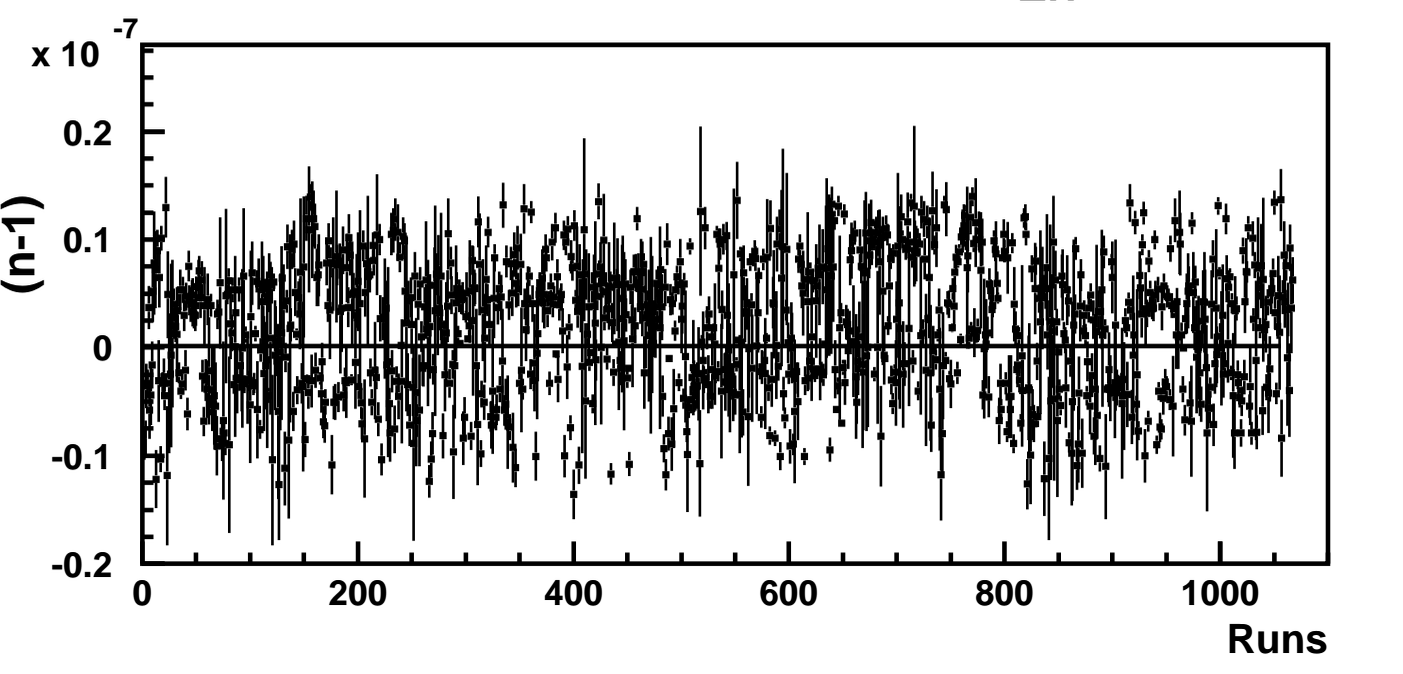} \includegraphics[scale=0.4]{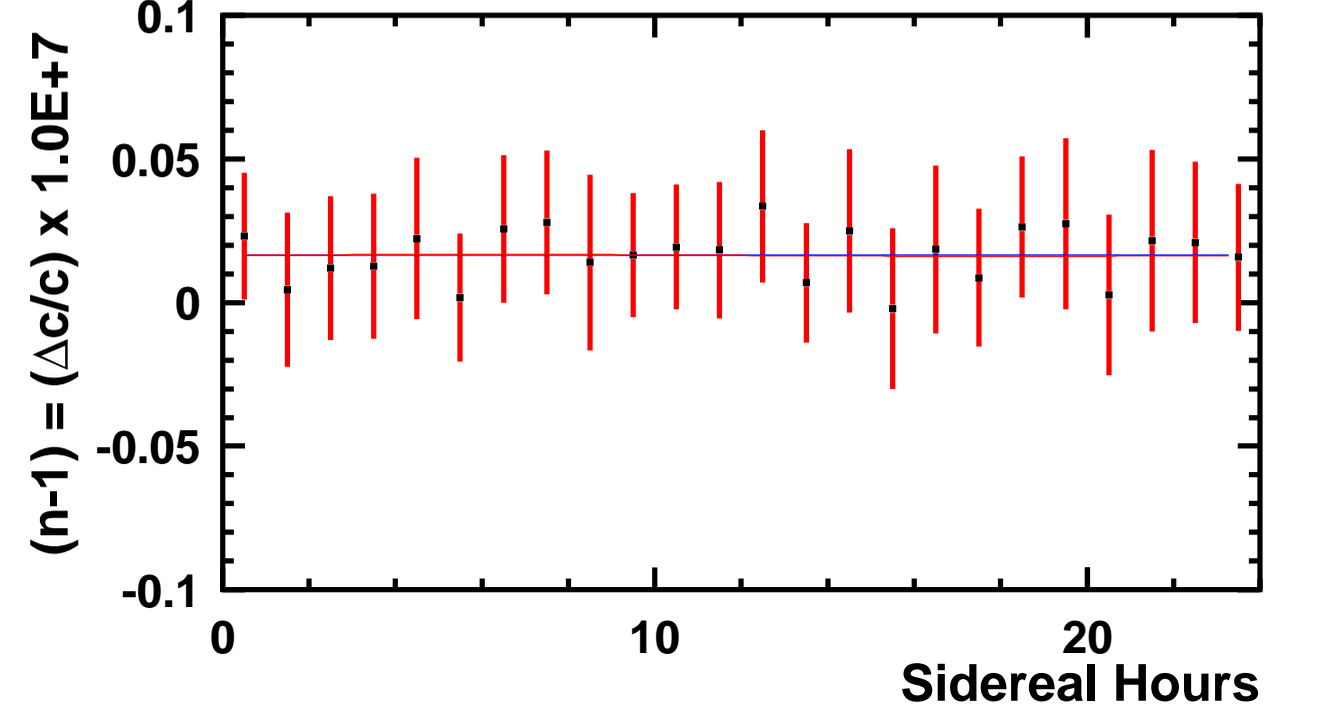} 
\caption[]{(Left) - Plot showing the variation of `$n-1$' as a function of simple time along with accrued errors \emph{i.e.} quadrature sum of statistical and systematic uncertainties. (Right) - Plot showing the variation of `$n-1$' as a function of sidereal time, with the entire data set from 1200 runs rolled into modulo sidereal day.}
\label{fig6}
\end{figure}

Using the asymmetry measured at on $\sim$ 50 strips of the electron detector in for the JLab Hall-C Compton polarimeter, one could perform a least squares fit to the expression for asymmetry w.r.t the photon energy derived above in order to obtain values of $n$. The data shown here are from the JLab Compton polarimeter which ran for approximately 6 months and the extracted values of $n$ are plotted w.r.t to run number (or data set) in FIG. \ref{fig6} (Left).

\subsection{Determination of limits on SME parameters}

MSME provides a number of ways to allow Lorentz and thus CPT violation \cite{[31]}. The leading MSME coefficient which causes direction and polarization dependent vacuum dispersion is $k_F$ \cite{[32]}. In this sub section, we will try extract the upper limit on the $\tilde{\kappa}_{0^+}$ components of $\kappa_F$, which is a $3 \times 3$ antisymmetric matrix. Using MSME, the dispersion relation for photon in terms of $\vec{\kappa}$ can be written as; $\omega = (1 - \vec{\kappa}.{\hat K})K + \mathcal{O}(\kappa^2)$ where $\vec{\kappa} = \left< \left( \tilde{\kappa}_{0^+}^{23} \right), \left( \tilde{\kappa}_{0^+}^{31} \right), \left( \tilde{\kappa}_{0^+}^{12} \right) \right> = \left<\kappa_X, \kappa_Y, \kappa_Z \right>$, $\omega$ is the energy of the photon and, $\vec{K}$ is the 3-momentum of the photon. Here Z-direction is parallel to the axes of rotation of the Earth. Using energy conservation in Compton scattering and the above relations, one could write the index of refraction of free space as;
\begin{eqnarray}
n &\approx& \left( 1 + \vec{\kappa} . \hat{p} \right) \approx 1 + \left( 0.87 \sqrt{\kappa_X^2 + \kappa_Y^2} sin(\Omega t) \right)
\label{eq:np}
\end{eqnarray}
where $\hat{p}(t)$ is unit vector along the 3-momentum of electron beam which for JLab Hall-C was $\left<0.13 cos(\Omega t), 0.87 sin(\Omega t), 0.48 \right>$, $K = 2.32$ eV is the momentum of the photon beam, $\left| \vec{p} \right| = \epsilon_0 = 1.157$GeV is the electron beam energy, $\gamma =  2280$ is the Lorentz boost of the electrons and $\Omega = 2\pi / (23h56m)$ is the frequency of rotation of the Earth. Finally Eq. \ref{eq:np} can be numerically expressed by disregarding the phase offset.

Furthermore, given that the JLab Compton polarimeter ran for almost 2 years, one could also study the variation of the refractive index of vacuum as a function of sidereal time and compare it to Eq. \ref{eq:np}. Fig. \ref{fig6} (right) shows a plot of this variation fit to a pure sinusoid wave with a frequency of $\Omega$, amplitude of ($0.3 \times 10^{-10} \pm 0.2 \times 10^{-9}$), and an offset of ($0.16 \times 10^{-8} \pm 0.02\times 10^{-8}$), with $\chi^2 = 1.24$ from $22$ d.o.f. With a value of sinusoidal fit amplitude being constrained at the $95\%$ confidence level to $< 0.4 \times 10^{-9}$, one could impose a limit on $\sqrt{\kappa_X^2 + \kappa_Y^2} <  4.6 \times 10^{-10}$ from Eq. \ref{eq:np}.

\section{Conclusion}
On averaging the values of $n$ obtained from the least squares fit to each data set in FIG. \ref{fig6} (Left), we obtain a value of $n = 1 + (2.44\times10^{-9} \pm 6.82 \times 10^{-9})$. While this is a zero result, a small non-zero value for vacuum dispersion could be explained by the effects of general relativity \cite{[33]}. A limit of $\sqrt{\kappa_X^2 + \kappa_Y^2} <  4.6 \times 10^{-10}$ also implies that $\sqrt{\left( 2c_{TX} - (\tilde{\kappa}_{0^+}^{YZ} \right)^2 + \left( 2c_{TY} - (\tilde{\kappa}_{0^+}^{ZX} \right)^2} < 4.6 \times 10^{-10}$ \cite{[4]}.

\end{document}